\documentclass[12pt]{article}

\usepackage{graphicx}

\newcommand{\tr}{\mbox{Tr} \, }
\newcommand{\ket}[1]{\left | #1 \right \rangle}
\newcommand{\bra}[1]{\left \langle #1 \right |}

\newcommand{\proj}[1]{\ket{#1} \! \bra{#1}}
\newcommand{\outerprod}[2]{\ket{#1} \! \bra{#2}}

\newcommand{\superop}{{\cal E}}

\newcommand{\relent}[2]{{\cal D}\left ( #1 || #2 \right )}
\newcommand{\supp}{\mbox{supp} \, }
\newcommand{\avail}{\mbox{$\cal A$}}
\newcommand{\bigsum}[1]{{\displaystyle \sum_{#1}}}

\begin{document}

\title{Optimal signal ensembles}
\author{Benjamin Schumacher$^{(1)}$
        and Michael D. Westmoreland$^{(2)}$}
\maketitle
\begin{center}
{\sl
$^{(1)}$Department of Physics, Kenyon College, Gambier, OH 43022 USA \\
$^{(2)}$Department of Mathematical Sciences, Denison University,
 Granville, OH  43023 USA }
\end{center}

\section*{Abstract}

Classical messages can be sent via a noisy quantum channel in
various ways, corresponding to various choices of ensembles of
signal states of the channel.  Previous work by Holevo and by 
Schumacher and Westmoreland relates the capacity of the channel
to the properties of the signal ensemble.  Here we describe some
properties characterizing the ensemble that maximizes the capcity,
using the relative entropy ``distance'' between density operators
to give the results a geometric flavor.

\section{Communication via quantum channels}

Suppose Alice wishes to send a (classical) message to Bob, using a
quantum system as the communication channel.  Alice prepares the
system in the ``signal state'' $\rho_k$ with probability $p_k$, so
that the ensemble of states is described by an average density
operator $\rho = \bigsum{k} p_k \, \rho_k$.
Bob makes a measurement of a ``decoding observable''
on the system and uses the result to infer which signal state
was prepared.  The choice of system preparation (represented
by the index $k$) and Bob's measurement outcome are the input
and output of a classical communication channel.

Holevo \cite{holevo1} proved (as Gordon \cite{gordon} and
Levitin \cite{levitin} had previously conjectured) that the
mutual information between the input and output of this channel,
regardless of Bob's choice of decoding observable, can never
be greater than $\chi$, where
\begin{equation}
        \chi = S(\rho) - \sum_{k} p_k S(\rho_k)   \label{chidef}
\end{equation}
where $S(\rho) = - \tr \rho \log \rho$ is the von Neumann
entropy of the density operator $\rho$.

More recently, it has been shown by Holevo \cite{holevo2} and by
Schumacher and Westmoreland \cite{noisy} that the Holevo bound is
asymptotically achievable.  That is, if Alice uses many copies of
the same channel, preparing long code words of signal states, and
if Bob chooses an entangled decoding observable, Alice can convey
to Bob up to $\chi$ bits of information per use of the channel,
with arbitrarily low probability of error.  (This fact was
first shown for pure state signals in \cite{hjsww}.)

Suppose the channel is a noisy one described by a superoperator
$\superop$.  Then if Alice prepares the input signal state
$\rho_k$, Bob will receive the output signal state
$\superop (\rho_k)$.  It is the ensemble of output
signal states that determines the capacity of the channel.
Effectively, the superoperator $\superop$ restricts the set
of signals that Alice can present to Bob for decoding.
If $\cal B$ is the set of all density operators, then
Alice's efforts can only produce output states in the
set $\avail = \superop ( {\cal B} )$.

In this paper we will consider the problem of maximizing $\chi$
for ensembles of states drawn from a given set
$\avail$ of available states.
This includes the problem of maximizing $\chi$ for the
outputs of a noisy channel, if $\avail$ is chosen to be the set
of possible channel outputs.  In this case, $\avail$ will be a
convex set; but we will not need the convexity of $\avail$ for
many of our results.

\section{Relative entropy}

If $\rho$ and $\sigma$ are density operators, then the
{\em relative entropy} of $\rho$ with respect to $\sigma$
is defined to be
\begin{equation}
\relent{\rho}{\sigma} = \tr \rho \log \rho - \tr \rho \log \sigma  .
                        \label{relentdef}
\end{equation}
Here are three important points about the relative entropy:
\begin{itemize}
\item  $\relent{\rho}{\sigma} \geq 0$, with equality if and
	only if $\rho = \sigma$.
\item  Strictly speaking, $\relent{\rho}{\sigma}$ is defined
	only if $\supp \rho \subseteq \supp \sigma$
	(where ``$\supp \rho$'' is the support of the operator
	$\rho$).  If this is not the case, then we take
	$\relent{\rho}{\sigma} = \infty$.  For
        example, if $\rho$ and $\sigma$ are distinct pure states,
        the relative entropy is always infinite.
\item  The relative entropy is jointly convex in its arguments:
        \begin{equation}
        \relent{p_{1} \rho_{1} + p_{2} \rho_{2}}{p_{1} \sigma_{1}
                + p_{2} \sigma_{2}}  \leq
                p_{1} \relent{\rho_{1}}{\sigma_{1}} +
                p_{2} \relent{\rho_{2}}{\sigma_{2}}
        \end{equation}
        for $p_{1}, p_{2} \geq 0$ with $p_{1} + p_{2} = 1$.
	From this fact it also follows that the relative entropy
	is convex in each of its arguments.
\end{itemize}
The relative entropy plays a role in the asymptotic distinguishability
of quantum states by measurement \cite{distinguish}, and has been used
to develop measures of quantum entanglement \cite{entangle}

It is often convenient to think of the relative entropy
$\relent{\rho}{\sigma}$ as a ``directed distance'' from $\sigma$
to $\rho$, even though it lacks some of the properties of a true metric.
This view of the relative entropy will let us give a geometric
interpretation to our results.

Suppose as before we have an ensemble of signal states in
the available set $\avail$, in which $\rho_k$ appears
with probability $p_k$.  It is easy to verify that the
Holevo bound $\chi$ can be given in terms of the relative entropy:
\begin{equation}
\chi     =   \sum_k p_k \relent{\rho_k}{\rho} .  \label{chirelent}
\end{equation}
That is, $\chi$ is just the average of the relative
entropy of the members of the signal ensemble with respect
to the average signal state.

\section{The optimal signal ensemble}

To maximize the information capacity of the channel,
Alice will want to choose a signal ensemble that maximizes $\chi$.
We will denote the maximum of $\chi$ for a given set
$\avail$ of available states by $\chi^{*}$.  Any ensemble
of signal states that achieves this value
of the Holevo bound will be called an {\em optimal} signal ensemble.

If the set of available states $\avail$ is a closed convex set,
then we can always take an optimal ensemble to be composed of
extreme points of $\avail$---that is, states which cannot be written
as convex sums of other states in $\avail$.
To see this, suppose we have an ensemble of $\avail$-states
with average state $\rho$, and further suppose that $\rho_{k}$
is a member of the ensemble that is not an extreme point.  This
means that there are states $\rho_{k0}$ and $\rho_{k1}$ in $\avail$
such that
\begin{equation}
        \rho_{k} = q_{0} \rho_{k0} + q_{1} \rho_{k1}
\end{equation}
for probabilities $q_{0}$ and $q_{1}$ that sum to unity.  By the
convexity
of the relative entropy,
\begin{equation}
        \relent{\rho_{k}}{\rho} \leq q_{0} \relent{\rho_{k0}}{\rho}
                + q_{1} \relent{\rho_{k1}}{\rho} .
\end{equation}
Since $\chi$ is the average of the relative entropies, we will never
make
$\chi$ smaller by replacing $\rho_{k}$ (with probability $p_{k}$) by
$\rho_{k0}$ and $\rho_{k1}$ (with probabilities $p_{k} q_{0}$ and
$p_{k} q_{1}$, respectively) in the ensemble.  Thus, at least
one optimal ensemble will be composed of extreme points of $\avail$.

For noisy channels, this means that pure state inputs to the channel
are optimal -- that is, it never increases $\chi$ to use
mixed states as inputs.  This fact was shown in \cite{noisy}.

A second and very surprising fact was discovered by Fuchs
\cite{fuchs}.  The quantity $\chi$ is a measure of
the distinguishability of an ensemble of signal states.
If we wish to maximize the distinguishability of the
output signals of a noisy channel, we might imagine that we should
always maximize the distinguishability
of the input signals---i.e., choose an orthogonal set
of input states.  But this intuition turns out to be false.

\begin{figure}
\begin{center}
\includegraphics[width=5in]{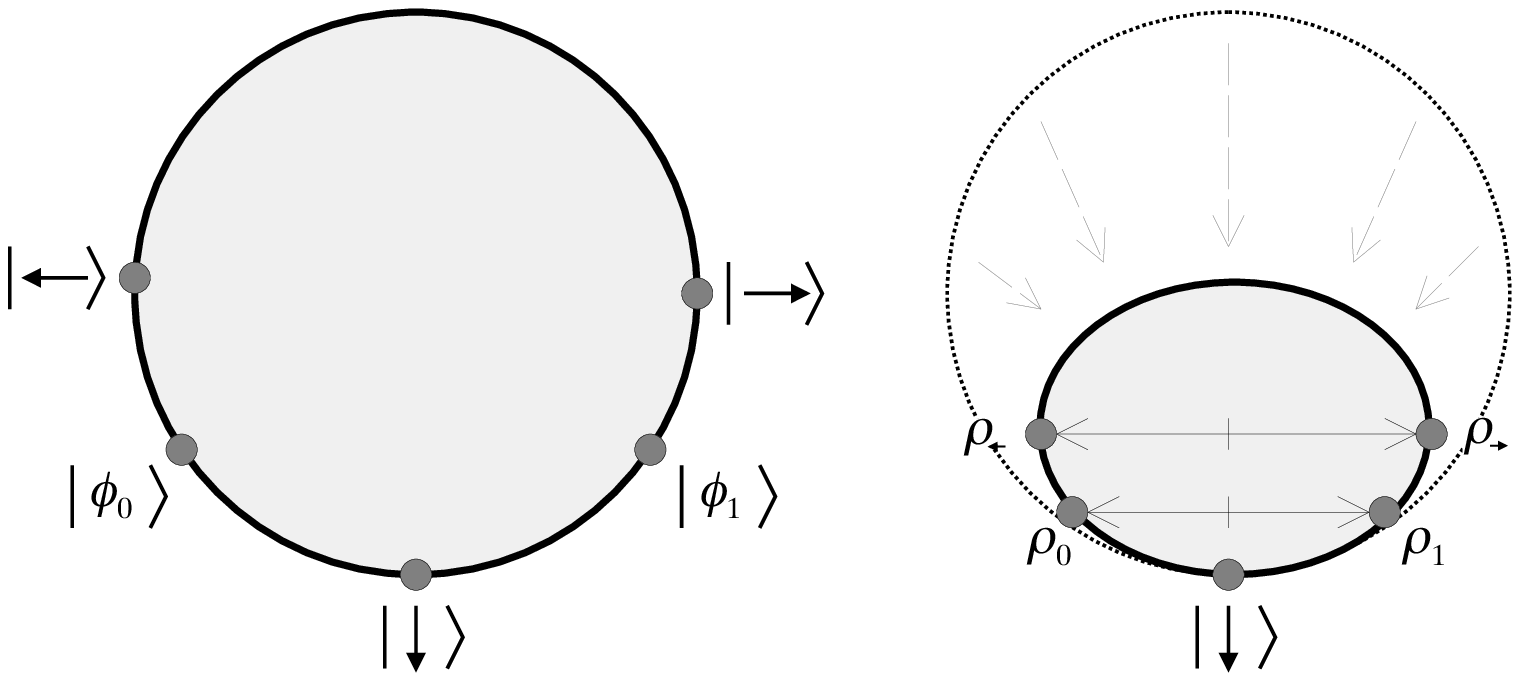}
\end{center}
\caption{Bloch sphere diagram for amplitude damping.
  The highest value of $\chi$ for a set of orthogonal
  input signals is attained by an equally weighted mixture
  of $\ket{\rightarrow}$ and $\ket{\leftarrow}$, but
  the non-orthogonal input signals $\ket{\phi_0}$
  and $\ket{\phi_1}$ yield
  a larger value of $\chi$.}
\label{fuchs}
\end{figure}

Some insight can be gained by examining a specific counter-example.
Our quantum system is a spin, and $\ket{\uparrow}$ and
$\ket{\downarrow}$
represent eigenstates of $S_{z}$. The spin is subject
to ``amplitude damping'',
so that an initial density operator $\rho$ evolves into a density
operator
\begin{equation}
        \rho' = \superop(\rho) = A_{1} \rho A_{1}^{\dagger}
                                + A_{2} \rho A_{2}^{\dagger}
\end{equation}
where $A_{1} = \sqrt{1-\lambda} \proj{\uparrow} + \proj{\downarrow}$
and
$A_{2} = \sqrt{\lambda} \outerprod{\downarrow}{\uparrow}$,
and $0 \leq \lambda \leq 1$.
The result of this operation is, for instance, to leave the state
$\ket{\downarrow}$ unchanged but to cause
$\ket{\uparrow}$ to decay to $\ket{\downarrow}$
with probability $\lambda$.  We choose $\lambda = 1/2$.
A diagram of
this process in the Bloch sphere is found in Figure~\ref{fuchs}.

If we consider only orthogonal input signal ensembles, the maximum
$\chi$ is obtained for an equally weighted ensemble of
$\ket{\rightarrow}$ and $\ket{\leftarrow}$,
for which $\chi = 0.4567$ bits.
But a non-orthogonal ensemble of the states
$\ket{\phi_0}$ and $\ket{\phi_1}$ can achieve
$0.4717$ bits, where the angle in
Hilbert space between the two inputs is about 80$^{\circ}$.

Why is this?  Recall that $\chi$ is the average relative entropy
``distance''
from the average signal state to the individual signal states.
This distance function grows
larger near the boundary of the Bloch
sphere--so that, for example, the relative entropy distance between
distinct pure states is infinite.  Thus, despite the appearance in
Figure~\ref{fuchs}, the relative entropy distances for the ensemble
of $\rho_{0}$ and $\rho_{1}$ are {\em greater} than those for the
ensemble of $\rho_{\rightarrow}$ and $\rho_{\leftarrow}$.

\section{Changing the ensemble}

In this section we will prove some useful results that
will enable us to further characterize the optimal ensembles
for a given set $\avail$ of available states.

Suppose as before that the signal state $\rho_{k} \in \avail$
appears in our ensemble with probability $p_{k}$,
yielding an average state $\rho$.
Let $\sigma$ be some other density operator,
which we will call the ``alternate'' state.
Then we can calculate the average relative entropy distance
of the signal states from $\sigma$:
\begin{eqnarray}
\sum_{k} p_k \relent{\rho_k}{\sigma}
        & = & \sum_{k} p_k \left ( \tr \rho_k \log \rho_k
                        - \tr \rho_k \log \sigma \right ) \nonumber \\
        & = & \sum_{k} p_k \left ( \tr \rho_k \log \rho_k
                        - \tr \rho_k \log \rho \right ) \nonumber \\
        &   & + \left ( \tr \rho \log \rho
                        - \tr \rho \log \sigma \right ) \nonumber \\
        &   &  \nonumber \\
        & = & \sum_k p_k \relent{\rho_k}{\rho} + \relent{\rho}{\sigma}
                \nonumber \\
\sum_{k} p_k \relent{\rho_k}{\sigma}
        & = & \chi + \relent{\rho}{\sigma} \label{coverish} .
\end{eqnarray}
This useful identity, first given by Donald\cite{donald},
has a number of implications.  For example,
\begin{itemize}
\item  For any ensemble and any $\sigma$,
        \begin{equation}
        \sum_k p_k \relent{\rho_k}{\sigma} \geq \chi
	\label{donaldcorollary}
        \end{equation}
        with equality if and only if $\sigma = \rho$.
\item  From the previous point it follows that
        \begin{equation}
        \chi = \min_{\sigma} \left ( \sum_k p_k
		\relent{\rho_k}{\sigma} \right )
        \end{equation}
        where the minimum is taken over all density operators
	$\sigma$.
\end{itemize}

Now we will use our identity to consider how the value of $\chi$
would change if we were to modify our ensemble.  In particular, we
can introduce a new state $\rho_{0}$ with probability $\eta$,
shrinking the other probabilities to maintain normalization.  We
may conveniently refer to our ensembles as the ``original'' and
``modified'' ensembles, as summarized in the following table:
\begin{center}
\begin{tabular}{lcc}
ensemble                &  original             &  modified  \\
signal states           &  $\rho_{k}$           &  $\rho_{k}, \rho_{0}$
\\
probabilities           &  $p_{k}$              &  $(1-\eta) p_{k},
\eta$ \\
average state           &  $\rho$               &  $\rho'$ \\
Holevo bound            &  $\chi$               &  $\chi'$
\end{tabular}
\end{center}
where
\begin{eqnarray}
\rho'  &  =  &  (1 - \eta) \rho + \eta \rho_{0}  \\
\chi  &  =  &  \sum_{k} p_{k} \relent{\rho_{k}}{\rho} \\
\chi'  &  =  &  (1-\eta)\sum_{k} p_{k} \relent{\rho_{k}}{\rho'} \,\, +
\,\,
                        \eta \relent{\rho_{0}}{\rho'} .
\end{eqnarray}
We wish to find how the Holevo bound changes -- that is, we wish to
make an estimate of $\Delta \chi = \chi' - \chi$.

Begin with the expression for $\chi'$ and apply
Equation~\ref{coverish}, choosing the original ensemble
and letting the modified average state
$\rho'$ play the role of the alternate state.  This yields
\begin{eqnarray*}
\chi'   & = &   (1 - \eta) \left ( \chi + \relent{\rho}{\rho'}
		\right )
                + \eta \relent{\rho_0}{\rho'}   \\
        & = &   \chi + \eta \left ( \relent{\rho_0}{\rho'}
		- \chi \right )
                + (1-\eta) \relent{\rho}{\rho'} \\
\Delta \chi
        & = & \eta \left ( \relent{\rho_0}{\rho'} - \chi \right )
                + (1-\eta) \relent{\rho}{\rho'} .
\end{eqnarray*}
Therefore,
\begin{equation}
\Delta \chi \geq \eta \left ( \relent{\rho_0}{\rho'} - \chi \right ) .
        \label{lower}
\end{equation}
This gives us a lower bound for $\Delta \chi$.

To obtain an upper bound, we apply Equation~\ref{coverish} to the
modified ensemble, with the original average state $\rho$ playing
the role of the alternate state.
\begin{eqnarray*}
\chi' + \relent{\rho'}{\rho}
        & = & (1-\eta) \left( \sum_{k} p_k \relent{\rho_k}{\rho}
		\right ) + \eta \relent{\rho_0}{\rho}  \\
        & = & (1-\eta) \chi + \eta \relent{\rho_0}{\rho} \\
\chi' - \chi & = & \eta \left ( \relent{\rho_0}{\rho}
		- \chi \right ) - \relent{\rho'}{\rho}
\end{eqnarray*}
And so we obtain
\begin{equation}
\Delta \chi \leq \eta \left ( \relent{\rho_0}{\rho}
		- \chi \right ) .
        \label{upper}
\end{equation}
In deriving this inequality, we obviously assume that
$\supp \rho_0 \subseteq \supp \rho$.  But if this
is not the case, then the inequality still holds
in the sense that the right-hand side is infinite.

It is easy to generalize these results to a situation in which
we modify the ensemble by adding many states.
Suppose the states $\rho_{0a}$ are added with probabilities
$\eta q_a$ (where the $q_a$'s form a probability distribution).
Then the above results would become
\begin{equation}
\eta \left( \sum_{k} q_a \relent{\rho_{0a}}{\rho'} - \chi \right )
        \leq \Delta \chi \leq
        \eta \left ( \sum_{k} q_a \relent{\rho_{0a}}{\rho}
		- \chi \right ).
        \label{manystate}
\end{equation}
All of our subsequent results still
hold in this more general situation,
but to simplify the discussion we will phrase our arguments
in terms of ``single state'' modifications of a given ensemble.

Finally, consider states $\rho_0$ and $\rho$, and let
$\rho' = (1-\eta) \rho + \eta \rho_0$.
Then $\relent{\rho_0}{\rho'}$ exists and is finite for
$0 < \eta \leq 1$,
and
\begin{itemize}
\item  If $\supp \rho_0 \subseteq \supp \rho$, then
        $\relent{\rho_0}{\rho'} \rightarrow \relent{\rho_0}{\rho}$
        as $\eta \rightarrow 0$.
\item  Otherwise, $\relent{\rho_0}{\rho'} \rightarrow \infty$ as
        $\eta \rightarrow 0$.
\end{itemize}
We see that Equations \ref{lower} and \ref{upper} are fairly
``tight'' lower and upper bounds for $\Delta \chi$,
because (informally speaking) the two expressions
approach one another as $\eta$ approaches zero.

\section{Properties of optimal ensembles}

For a given set ${\cal A}$ of available states (e.g., the
outputs of a noisy channel),
let $\rho_k$ and $p_k$ be the members and probabilities
of the ensemble of ${\cal A}$-states for which $\chi$ takes
on its maximum value.  Call this the ``$\chi$-optimal ensemble'',
and let $\rho^{\ast}$ be the average state of this ensemble.
Denote $\max \chi$ by $\chi^{\ast}$.
The $\chi$-optimal ensemble has a number
of important properties.

\noindent
\begin{description}
\item[Existence.] If the letter states are outputs of a
noisy channel in a finite-dimensional Hilbert space,
then a $\chi$-maximizing ensemble exists.

  {\bf Proof:} The key result can found in \cite{uhlmann}:
Let $\avail$ be a convex, compact subset
$\avail$ of density operators
on a Hilbert space of finite dimension $d$, and let
$\rho$ be in $\avail$.
If the set of extremal elements of $A$
is compact then for any $\rho \in A$ there
exists an ensemble of states $\{ \rho_k \} \subset A$
with $\rho = \sum p_k \rho_k$
that maximizes $\chi$ over the set
of all ensembles whose average state is $\rho$.
In other words, there exist optimal signal ensembles
for a given average state $\rho$.
By Caratheodory's Theorem, since the Hilbert space
has $d$ dimensions, then there are optimal ensembles
(in this sense) with no more than $d^2$ states.

We see that the conditions for the result from \cite{uhlmann}
are met. The set of states $\avail$ that are possible outputs
of the channel is a convex, compact set with a compact set of
extremal points.  For any average state $\rho$ in $\avail$,
we can find a $\rho$-fixed optimal ensemble with $d^2$ or
fewer elements.
Thus, in order to maximize $\chi$ over all possible ensembles,
we only need to consider  the set of ensembles with no more
than $d^2$ elements drawn from $\avail$.  As this is a
finite cartesian product of a compact set, it is compact.
As $\chi$ is a continuous function, it must achieve its
maximum in this set of ensembles.  Thus, the existence of
an optimal ensemble of states in $\avail$ is assured.

\item[Maximal distance property.]  For any state $\rho_0$
	in ${\cal A}$,
        \begin{equation}
        \relent{\rho_0}{\rho^{\ast}} \leq \chi^{\ast} .
        \end{equation}

{\bf Proof.}  We assume the existence of a state $\rho_0$
with $\relent{\rho_0}{\rho^{\ast}} > \chi^{\ast}$.
(We allow for the possibility that
$\relent{\rho_0}{\rho^{\ast}}$
is infinite.)  Since $\relent{\rho_0}{\rho'} \rightarrow
\relent{\rho_0}{\rho^{\ast}}$ as $\eta \rightarrow 0$,
we can find a value of $\eta$ so that $\relent{\rho_0}{\rho'} >
\chi^{\ast}$.  Then by Equation~\ref{lower},
\begin{displaymath}
        \Delta \chi \geq
        \eta \left ( \relent{\rho_0}{\rho'} - \chi^{\ast} \right )
                > 0 .
\end{displaymath}
That is, we can increase $\chi$ by including $\rho_0$ in the
signal ensemble, which is a contradiction.
\item[Maximal support property.]  For a $\chi$-optimal ensemble,
        $\supp \rho^{\ast} = \supp {\cal A}$.
	(By ``$\supp {\cal A}$'' we mean the smallest
	subspace that contains $\supp \rho_k$ for any
	$\rho_k \in {\cal A}$.)  In other words, any
        $\chi$-optimal ensemble ``covers'' the support of the set
        of available states.

{\bf Proof.}  This is a corollary to the maximum distance
property.  If there were a state $\rho_{0} \in {\cal A}$
so that $\supp \rho_0$ were not contained in $\supp
\rho^{\ast}$,  then $\relent{\rho_0}{\rho^{\ast}}$
would be infinite.

\item[Sufficiency of maximal distance property.]  Suppose
	we have an ensemble with average state $\rho$ and
	a particular value of $\chi$, and suppose that
        \begin{displaymath}
        \relent{\rho_0}{\rho} \leq \chi
        \end{displaymath}
        for all $\rho_0 \in {\cal A}$.  Then this must be a
	$\chi$-optimal  ensemble.
	That is, the only ensembles that have the maximal
        distance property are $\chi$-optimal ensembles.

{\bf Proof:}  If we add a state $\rho_0$ with probability $\eta$
to the ensemble, then from Equation~\ref{upper}
        \begin{displaymath}
        \Delta \chi \leq \eta \left ( \relent{\rho_0}{\rho}
		- \chi \right )  \leq 0
        \end{displaymath}
so that we cannot increase $\chi$.  (By Equation~\ref{manystate},
the same would hold if we were to add several
different states instead of only one.)  Thus,
$\chi = \chi^{\ast}$.

\item[Equal distance property.]  Suppose $\rho_k$ is a member
	of a $\chi$-optimal ensemble with probability
	$p_k \neq 0$.  Then
        \begin{equation}
        \relent{\rho_k}{\rho^{\ast}} = \chi^{\ast} .
        \end{equation}
        In other words, all of the non-zero members of a $\chi$-optimal
        ensemble have the same relative entropy ``distance'' with
        respect to the average state $\rho^{\ast}$.

{\bf Proof:}  This is another corollary to the maximal distance
property.  If $\relent{\rho_k}{\rho^{\ast}} < \chi^{\ast}$
for any $\rho_k$ with $p_k \neq 0$, then the average relative
entropy cannot equal $\chi^{\ast}$.

\item[Min-max formula for $\chi^{\ast}$.]  From the above
	properties, we can show the following formula:
        \begin{equation}
        \chi^{\ast} = \min_{\rho} \left (
                        \max_{\rho_0} \relent{\rho_{0}}{\rho}
			\right ) ,
		\label{minmax}
        \end{equation}
        where the maximum is taken over all
	$\rho_{0} \in {\cal A}$ and the minimum is taken over
	all average states $\rho$ of ensembles
        of ${\cal A}$-states.

{\bf Proof:}  We first show that, for any state $\sigma$, the
quantity $\displaystyle \max_{\rho_0} \relent{\rho_{0}}{\sigma}$
is an upper bound for the value of $\chi$ for any possible
ensemble.  By Equation~\ref{donaldcorollary}, we find that
\begin{displaymath}
	\chi \leq \sum_{k} p_{k} \relent{\rho_{k}}{\sigma}
		\leq \max_{\rho_0} \relent{\rho_{0}}{\sigma} .
\end{displaymath}
This will also hold for an optimal signal ensemble,
for which $\chi = \chi^{\ast}$.  Thus,
\begin{displaymath}
	\chi^{\ast} \leq \min_{\rho} \left (
                        \max_{\rho_0} \relent{\rho_{0}}{\rho}
			\right ) .
\end{displaymath}
Next we note that the maximal distance property implies that
\begin{displaymath}
	\chi^{\ast} = \max_{\rho_0} \relent{\rho_{0}}{\rho^{\ast}} ,
\end{displaymath}
from which we can see that
\begin{displaymath}
	\chi^{\ast} \geq \min_{\rho} \left (
                        \max_{\rho_0} \relent{\rho_{0}}{\rho}
			\right ) .
\end{displaymath}
These two inequalities establish the formula in Equation~\ref{minmax}.

\end{description}

These properties provide strong characterizations of an
optimal signal ensemble for a quantum channel.  Equation~\ref{minmax},
for example, shows that $\chi^{\ast}$ can be calculated as a purely
``geometric'' property of the set $\avail$, without direct reference
to any ensemble.  We believe that our results are likely to prove
useful in further investigations of the efficient use
of quantum resources to transmit classical messages.

\section{Acknowledgements}

We would like to thank A. Uhlmann for helpful and enlightening
comments, particularly about the existence of an
optimal ensemble.  We also had useful conversations with
T. Cover, C. A. Fuchs, A. S. Holevo, V. Vedral
and W. K. Wootters.  Most of these discussions took place in
connection with the programme on ``Complexity, Computation
and the Physics of Information'' at the Isaac Newton Institute
in Cambridge (England) during the summer of 1999.
This programme was sponsored in part by the European
Science Foundation.  One of us (BS) gratefully acknowledges
the support of a Rosenbaum Fellowship at the Isaac Newton
Institute to participate in this programme.

\newpage

\end{document}